\documentclass[letter]{aa} % for the letters 
\usepackage{graphicx}
\usepackage{txfonts}
\usepackage{natbib}
\usepackage[breaklinks=false]{hyperref}
\newcommand{\dif}{{\rm d}}

\newcommand{\Ms}{{\ensuremath{\mathrm{M}_{\odot}}}}
\newcommand{\Rs}{{\ensuremath{\mathrm{R}_{\odot}}}}

\newcommand{\Mpy}{\Ms\,{\rm yr}{\ensuremath{^{-1}}}}
\newcommand{\dm}{\ensuremath{\dot M}}
\newcommand{\gva}{{\sc genec}}
\newcommand{\vit}{{\ensuremath{v}}}
\newcommand\cst{\ensuremath{{\rm cst}}}

\newcommand{\dmax}{\ensuremath{\dot M_{\rm max}}}

\newcommand{\tff}{{\ensuremath{t_{\rm FF}}}}
\newcommand{\vff}{{\ensuremath{\vit_{\rm FF}}}}

\newcommand{\vvvff}{{\ensuremath{\vit^3_{\rm FF}}}}
\newcommand{\vson}{{\ensuremath{v_{s}}}}

\newcommand{\vvvson}{{\ensuremath{v^3_{s}}}}
\newcommand{\robar}{{\ensuremath{\bar\rho}}}

\usepackage{color}

\begin{document}

\title{On the maximum accretion rate of supermassive stars}
%\titlerunning{.}

\author{
L. Haemmerl\'e\inst{\ref{inst1}},
R. S. Klessen\inst{\ref{inst2},\ref{inst3}},
L. Mayer\inst{\ref{inst4}},
L. Zwick\inst{\ref{inst4}}
}
\authorrunning{Haemmerl\'e et al.}

\institute{
D\'epartement d'Astronomie, Universit\'e de Gen\`eve, chemin des Maillettes 51, CH-1290 Versoix, Switzerland        \label{inst1}\and
Universit\"at Heidelberg, Zentrum f\"ur Astronomie, Institut f\"ur Theoretische Astrophysik, Albert-Ueberle-Str. 2, D-69120 Heidelberg,
Germany \label{inst2}\and
Universit\"{a}t Heidelberg, Interdisziplin\"{a}res Zentrum f\"{u}r Wissenschaftliches Rechnen, Im Neuenheimer Feld 205, D-69120 Heidelberg,
Germany	\label{inst3}\and
Center for Theoretical Astrophysics and Cosmology, Institute for Computational Science, University of Zurich, Winterthurerstrasse 190,
CH-8057 Zurich, Switzerland     \label{inst4}
}

%\date{Received ; accepted }

% \abstract{}{}{}{}{} 
% 5 {} token are mandatory
 
\abstract
{The formation of the most massive quasars observed at high redshifts requires extreme inflows of gas down to the length scales of the central compact object.}
{Here, we estimate the maximum inflow rate allowed by gravity down to the surface of supermassive stars,
the possible progenitors of these supermassive black holes.}
{We use the continuity equation and the assumption of free-fall to derive maximum allowed inflow rates for various density profiles.
We apply our approach to the mass-radius relation of rapidly accreting supermassive stars
to estimate an upper limit to the accretion rates allowed during the formation of these objects.}
{We find that the maximum allowed rate \dmax\ is given uniquely by the compactness of the accretor.
For the compactness of rapidly accreting supermassive stars, \dmax\ is related to the stellar mass $M$ by a power-law $\dmax\propto M^{3/4}$.
The rates of atomically cooled halos (0.1 -- 10 \Mpy) are allowed as soon as $M\gtrsim1$ \Ms.
The largest rates expected in galaxy mergers ($10^4-10^5$ \Mpy) become accessible once the accretor is supermassive ($M\gtrsim10^4$ \Ms).}
{These results suggest that supermassive stars can accrete up to masses $>10^6$ \Ms\ before they collapse via the general-relativistic instability.
At such masses, the collapse is expected to lead to the direct formation of a supermassive black hole even within metal-rich gas, 
resulting in a black hole seed that is significantly heavier than in
conventional direct collapse models for atomic cooling halos.}

   %\keywords{}
 
\maketitle
%
%________________________________________________________________

\section{Introduction}
\label{sec-in}

The existence of quasars at redshift $\sim7$, hosting supermassive black holes (SMBHs) with masses $\gtrsim10^9$ \Ms\
\citep{mortlock2011,banados2018,wang2018,yang2020,wang2021},
implies extreme inflow rates down to small length-scales at the centre of the host galaxy during the early stages of galaxy formation.
The age of the Universe at redshift 7 is about half a billion years, so that these black holes must have gained mass at an average rate exceeding $\sim1$ \Mpy.
Only few astrophysical scenarios can lead to such high accretion rates.
In scenarios relying on seed black holes of only tens to hundreds of solar masses formed by the collapse of metal-free Population III stars at $z > 20$
this can hardly be sustained due to the ionizing bubbles that easily stifle accretion \citep{haemmerle2020a}.
In direct collapse scenarios a significantly more massive black hole seed results from the collapse of a precursor object assembled via
high inflow rates (e.g.~\citealt{woods2019,haemmerle2020a}).
The most studied  of these scenarios relies on inflows driven by the gravitational collapse of protogalaxies
hosted in primordial, atomically cooled haloes, reaching temperatures $\sim10^4$ K
(e.g.~\citealt{bromm2003b,dijkstra2008,latif2013e,regan2016a,regan2017})
at which the ratio of the Jeans mass to the free-fall time gives typical inflow rates of 1 \Mpy.
These rates are found down to  $0.01 - 0.1$ pc in hydrodynamical simulations \citep{latif2013e,chon2018,patrick2020}.
This scenario requires efficient dissociation of molecular hydrogen to avoid widespread
fragmentation that would suppress the gas inflow, which in turn demands
rather special environmental conditions \citep{woods2019}.
More recently, another scenario has been found that does not need the same restrictive conditions.
Hydrodynamical simulations have shown that the merger of massive, gas-rich galaxies at redshifts 8 -- 10,
can trigger gas inflow as high as $10^4-10^5$ \Mpy\ down to the resolution limit of 0.1 pc
\citep{mayer2010,mayer2015,mayer2019}.

The formation of the black hole seed represents a critical step in the formation of the earliest SMBHs
(e.g.~\citealt{woods2019,haemmerle2020a}).
The most important parameter is its mass, which is key for the efficiency of further accretion
(e.g.~\citealt{rees1978,rees1984,volonteri2010,volonteri2010b,valiante2017,zhu2020}).
It also determines the possible observational signatures of SMBH formation \citep{liu2007,shibata2016b,uchida2017,sun2017,sun2018,li2018}.
The direct progenitor of SMBHs in the scenarios mentioned above is thought to be a supermassive star (SMS),
growing by accretion up to masses $\gtrsim10^5$ \Ms\
\citep{hosokawa2012a,hosokawa2013,sakurai2015,umeda2016,woods2017,haemmerle2018a,haemmerle2018b,haemmerle2019c}
and eventually collapsing through the general-relativistic (GR) instability \citep{chandrasekhar1964}.
Its mass at collapse depends sensitively on the accretion rate \citep{haemmerle2020c,haemmerle2021a,haemmerle2021b}.
In particular, rates $\gtrsim100-1000$ \Mpy\ are required for final masses $\gtrsim10^6$ \Ms.
This mass-threshold allows to avoid supernova explosion even with a significant abundance of metals \citep{montero2012},
which might be present in massive galaxies at redshifts 8 -- 10.

The hydrodynamical simulations of atomically cooled haloes and galaxy mergers are limited to a resolution of 0.01 -- 0.1 pc.
In contrast, the radius of rapidly accreting SMSs is expected to be of the order of hundreds of AU
\citep{hosokawa2012a,hosokawa2013,schleicher2013,haemmerle2018a},
which is 1 -- 2 orders of magnitude smaller.
The detailed properties of the accretion flow in the gap between the resolution limit and the accretion shock
rely on complexe non-axisymmetric hydrodynamics, and depend on the thermal properties of the inflowing gas.
However, we show in the present work that an upper limit for the accretion rate, given uniquely by the compactness of the accretor,
arises from the fact that purely gravitational collapse cannot exceed free-fall, and proceeds always from lower to higher densities regions.
When combined with the mass-radius relation of rapidly accreting SMSs, this maximum accretion rate is given uniquely by the mass of the SMS,
and implies an upper limit for this mass, as a consequence of the GR instability.
The method relies on a simple application of the equation of continuity, and is detailed in section~\ref{sec-meth}.
The numerical estimates of the maximum accretion rates are presented in section~\ref{sec-res}.
The implications of these results are discussed in section~\ref{sec-dis},
and we summarise our conclusions in section~\ref{sec-out}.

\section{Method}
\label{sec-meth}

According to the equation of continuity, the spherical inflow of mass in a sphere of radius $r$
is given by the density of the material $\rho$ and its inwards radial velocity \vit\ evaluated at $r$:
\begin{equation}
\dm:=\left.{\dif M_r\over\dif t}\right\vert_r=4\pi r^2\rho\vit={3\vvvff\over2G}{\rho\over\robar}{\vit\over\vff}\;,
\label{eq-dmin}\end{equation}
where $M_r$ is the mass enclosed inside $r$.
In equation~(\ref{eq-dmin}), we have expressed the density and the velocity in dimensionless form,
as fractions of the average density inside $r$ and the free-fall velocity in $r$:
\begin{eqnarray}
\robar&=&{M_r\over{4\over3}\pi r^3}	\label{eq-robar} \;,\\
\vff&=&\sqrt{2GM_r\over r}			\label{eq-vff}\;,
\end{eqnarray}
where $G$ is the gravitational constant.
The free-fall velocity \vff\ is the speed that the layer with mass-coordinate $M_r$ would have at $r$ if it collapsed from infinity without facing a pressure gradient.
Thus, it is the maximum velocity that pure gravitational attraction can produce.
It is given uniquely by the compactness of the mass $M_r$, defined by
\begin{eqnarray}
{r_S\over r}={2GM_r\over rc^2},
\label{eq-rsr}\end{eqnarray}
where $r_S=2GM_r/c^2$ is the local Schwarzschild radius and $c$ the speed of light.

Equation~(\ref{eq-dmin}) shows that, without density inversion ($\rho\leq\robar$), the inflow rate triggered by gravity ($\vit\leq\vff$) has an upper limit,
given uniquely by the free-fall velocity (\ref{eq-vff}) at radius $r$, that is by the compactness (\ref{eq-rsr}) of the mass $M_r$:
\begin{eqnarray}
\dm\leq\dmax:={3\vvvff\over2G}&=&{3c^3\over2G}\left({r_S\over r}\right)^{3/2}	\label{eq-dmax}\\
&=&0.96\times10^{13}\ \Mpy\left({r_S\over r}\right)^{3/2}	\;.				\label{eq-dmlim}
\end{eqnarray}
Only density inversion, large enough to ensure $\rho>\robar$, would allow to exceed this limit.
This condition requires more than local density inversion $\dif\rho/\dif r>0$,
and is inconsistent with gravitational collapse, in which matter flows from low-density to high-density regions.
This condition might be met only with an external trigger, such as an implosion or highly supersonic convergent flows.

The maximum rate (\ref{eq-dmax}) is reached for free-fall with homogeneous density $\rho=\cst$.
This last assumption is easily replaced by any power-law
\begin{equation}
\rho\propto r^{-\alpha}
\label{eq-rho}\end{equation}
with $\alpha<3$.
The limit $\alpha\to3$ corresponds to infinite mass-concentration $M_r=\cst$.
The models of isothermal collapse suggest density profiles with $\alpha\simeq2$
\citep{larson1969,penston1969a,shu1977,whitworth1985,maclow2004}.
For given $\alpha$, the ratio $\rho/\robar$ is constant:
\begin{equation}
{\rho\over\robar}=1-{\alpha\over3}\;.
\label{eq-alpha}\end{equation}
Inserting (\ref{eq-alpha}) into (\ref{eq-dmin}), we find a free-fall rate ($\vit=\vff$) of
\begin{equation}
\dm={3\vvvff\over2G}{\rho\over\robar}=\left(1-{\alpha\over3}\right){3\vvvff\over2G}\;.
\label{eq-dmalpha}\end{equation}
In the limit $\alpha\to3$, the rate goes to zero, reflecting the infinitesimal amount of mass near $r$ compared to the central mass for such centralised distribution.
For $\alpha=2$, we obtain a correction factor of 1/3 compared to the homogeneous case:
\begin{equation}
\dm={\vvvff\over2G}={\dmax\over3}.
\label{eq-dm2}\end{equation}

The maximum accretion rates of equations (\ref{eq-dmax}) or (\ref{eq-dmalpha})
are given uniquely by the compactness of the accretor, that is by its mass and its radius (equation~\ref{eq-rsr}).
It can be reduced to a $M-\dm$ relation, provided we also have a mass-radius relation for the accretor.
The mass-radius relation of rapidly accreting SMSs can be approximated by a power-law \citep{hosokawa2012a},
\begin{eqnarray}
R=260\,\Rs\left({M\over\Ms}\right)^{1/2},
\label{eq-rsms}\end{eqnarray}
which well reproduces the numerical models for any accretion rate $\dm\gtrsim0.01$ \Mpy\ and for masses $\lesssim10^5$ \Ms\
\citep{hosokawa2012a,hosokawa2013,haemmerle2018a,haemmerle2019c}.
The fact that the mass-radius relation reduces to a unique power-law, independent of the accretion rate,
results from the evolution along the Eddington and Hayashi limits.
High accretion rates favour large radii, so that rapidly accreting SMSs evolve as 'red supergiant protostars' \citep{hosokawa2013},
with nearly constant effective temperature.
On the other hand, the luminosity of SMSs is always close to the Eddington limit,
\begin{equation}
L\simeq L_{\rm Edd}={4\pi cGM\over\kappa}\;,
\end{equation}
where $\kappa=\cst$ is the opacity, dominated by electron scattering.
By definition, the effective temperature is related to the luminosity and the radius by
\begin{equation}
L=4\pi R^2\sigma T_{\rm eff}^4\;,
\end{equation}
where $\sigma$ is the Stefan-Boltzmann constant.
Thus, a constant $T_{\rm eff}$ and $L\propto M$ implies $R\propto M^{1/2}$.
Relation~(\ref{eq-rsms}) is found for $\kappa=0.35$ cm$^2$ g$^{-1}$ and $T_{\rm eff}=5000$ K, typical for red supergiants.
Inserting this relation  with $r=R$ and $M_r=M$ into expressions (\ref{eq-dmax}) or (\ref{eq-dmalpha}),
we obtain the maximum accretion rates of SMSs consistent with gravity, that of free-fall,
as a function of the mass of the accretor and for different density profiles.

\section{Results}
\label{sec-res}

\begin{figure}\begin{center}
\includegraphics[width=0.49\textwidth]{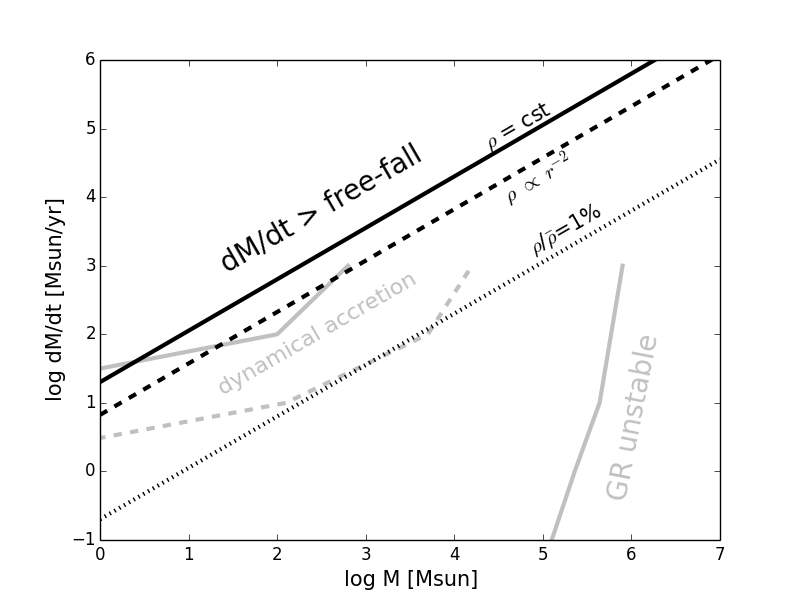}
\caption{Maximum accretion rates of SMSs following the mass-radius relation~(\ref{eq-rsms}), as a function of their mass,
for different ratios $\rho/\robar$ in the accretion flow (solid, dashed and dotted black lines).
The limits to hydrostatic equilibrium are shown in grey:
the limits of dynamical accretion \citep{haemmerle2019c}, indicate the maximum rates below which equilibrium is ensured by sound-waves,
in the core only (solid line) and in the whole star (dashed line);
the limit of stability against GR corrections is shown for the non-rotating case \citep{haemmerle2020c,haemmerle2021a}.}
\label{fig-mmdot}
\end{center}\end{figure}

The mass-radius relation (\ref{eq-rsms}) gives the following compactness for maximally accreting SMSs:
\begin{equation}
{R_S\over R}={2GM\over Rc^2}
=1.6\times10^{-6}\left({M\over10^4\,\Ms}\right)^{1/2}\;.
\label{eq-rsrsms}\end{equation}
Inserting (\ref{eq-rsrsms}) into (\ref{eq-dmax}), we obtain
\begin{equation}
\dmax=2\times10^4\ \Mpy\left({M\over10^4\,\Ms}\right)^{3/4}\;.
\label{eq-dmsms}\end{equation}
This limit is shown as a solid black line in figure~\ref{fig-mmdot}.
The maximum rate obtained for a density profile with $\alpha=2$ (equation~\ref{eq-dm2}) is shown as a black dashed line,
and that for $\rho/\robar=1\%$ (equation~\ref{eq-dmalpha}) as a black dotted line.
A profile with $\alpha=2$ reduces the rate by 1/3 only compared to the homogeneous case, following equation (\ref{eq-dm2}).
Obviously, the rate is decreased by 2 orders of magnitude for $\rho/\robar=1\%$.
We see that rates $\gtrsim100$~\Mpy\ always need masses $\gtrsim10$~\Ms, even in the most favourable situation of a homogeneous density profile.
For very steep density profiles, the minimum mass rapidly increases, reaching $\sim10^4$ \Ms\ for $\rho/\robar=1\%$.
Rates as high as $10^4-10^5$ \Mpy\ require a supermassive accretor even for homogeneous density,
whereas for $\rho/\robar=1\%$ the minimum mass exceeds $10^6$ \Ms.

\section{Discussion}
\label{sec-dis}

\subsection{Impact of the mass-radius relation}

While the maximum rate of expression~(\ref{eq-dmax}), as a function of the compactness (\ref{eq-rsr}),
relies only on the assumptions $\vit\ll\vff$ and $\rho\leq\robar$,
the maximum rate (\ref{eq-dmsms}), expressed as a function of the mass only, depends also on the choice of the mass-radius relation (\ref{eq-rsms}).
This assumption represents the main caveat of our approach.
The models of rapidly accreting SMSs \citep{hosokawa2012a,hosokawa2013,haemmerle2018a,haemmerle2019c},
that reproduce well this mass-radius relation for $M\lesssim10^5$ \Ms,
are based on several assumptions that might not be satisfied in the case of extreme accretion.
In particular, the ram pressure from the accretion flow, which is neglected in these models, might lead to compressed, more compact SMSs.
As a consequence of their higher compactness, such SMSs could accrete at larger rates than those following relation~(\ref{eq-rsms}).
From equation~(\ref{eq-dmax}), the maximum rate scales with $(M/R)^{3/2}$,
so that allowing for a rate 10 times higher requires a compression by a factor $\sim5$.
This is a significant factor but, because of the large density contrasts in SMSs, it requires only the outer percent of the stellar mass to be compressed.
The study of such compressed SMSs is beyond the scope of the present work,
but the results of section~\ref{sec-res} show already that
rates above the limit of equation~(\ref{eq-dmsms}) might only be reached by a new class of SMSs, while
rates below this limit remain consistent with existing models.

\subsection{Accretion rates of direct collapse}

Two channels of direct collapse have been studied in the literature:
the gravitational collapse of primordial, atomically cooled halos (e.g.~\citealt{bromm2003b,dijkstra2008,latif2013e,regan2016a,regan2017})
and the merger of two equal-mass gas-rich galaxies \citep{mayer2010,mayer2015,mayer2019}.
In the first scenario, metal-free gas is found to collapse with mass inflow rates 0.1 -- 10 \Mpy\ down to at least 0.01 -- 0.1 pc \citep{latif2013e,chon2018,patrick2020}.
The collapse proceeds nearly isothermally for most of the halo, which results in density profiles of $\rho\sim r^{-2}$.
But in the inner $\sim10^5$ \Ms\ the density profile is found to be approximately flat \citep{latif2013e}.
In this case, the inflows of 0.1 -- 10 \Mpy\ appear fully consistent with accretion down to the radius of a SMS.
Indeed, figure~\ref{fig-mmdot} shows that for both profiles, flat or $\rho\sim r^{-2}$, these rates are consistent with free-fall as soon as $M\gtrsim1$ \Ms.

Much larger rates are found in the case of galaxy mergers.
The deep potential well resulting from the merger leads to the formation of a supermassive disc of $\sim10^9$ \Ms,
that accretes at rates $10^4-10^5$ \Mpy\ \citep{mayer2010,mayer2015,mayer2019}.
The disc has a radius of a fraction of a parsec, which corresponds to a compactness of $\sim10^{-4}$.
Inserting this value into equation~(\ref{eq-dmax}), we obtain a maximum allowed accretion rate of $\sim10^7$ \Mpy\ for homogeneous density.
The hydrodynamical simulations show local fluctuations in the density profiles, due to instabilities and inhomogeneities in the disc,
but on average it follow the relation $\rho\sim r^{-2}$ down to the resolution limit of 0.1 pc.
As seen in equation~(\ref{eq-dm2}), the maximum rates in this case are decreased by a factor of 3 compared to the homogeneous case,
i.e. it remains in the same order of magnitude.
We see that the rates found in the hydrodynamical simulations, ranging up to $10^5$~\Mpy,
represent only a percent of the maximum accretion rate allowed by gravity.
Equation~(\ref{eq-dmax}) indicates that a compactness $r_S/r\gtrsim10^{-6}$ is a prerequesite for such inflows.
A star that obeys the mass-radius relation~(\ref{eq-rsms}) must be already supermassive ($\gtrsim10^4-10^5$ \Ms)
in order to accrete at rates $\sim10^4-10^5$ \Mpy\ (equation~\ref{eq-dmsms}).
On the other hand, rates above the upper limit for atomically cooled halos ($\gtrsim10-100$ \Mpy) are consistent with pure gravitational attraction
as soon as the mass of the accretor is $\gtrsim10$ \Ms.

\subsection{Maximally accreting SMSs}

The properties of stars accreting at rates $\geq100$ \Mpy\ have been studied in \cite{haemmerle2019c}.
For such accretion rates, the evolutionary timescales are so short that entropy losses remain negligible, and the evolution is governed by accretion.
The star contracts adiabatically, as a result of the mass increase at the surface, which implies a growing pressure in the centre.
The response of the core to the changes at the surface requires the pressure excess to be communicated inwards by sound waves.
If the characteristic time for accretion is shorter than the sound-crossing time, hydrostatic equilibrium is not ensured.
The evolution proceeds dynamically, and depends on the properties of the accretion flow.
We call this regime 'dynamical accretion'.
In \cite{haemmerle2019c}, we estimate the limit of this regime on the basis of hydrostatic stellar models at constant accretion rates,
corrected by accounting for the finite sound speed.

The limits of dynamical accretion in the whole star and in the envelope only are shown in figure~\ref{fig-mmdot} as solid and dashed grey lines, respectively.
An object of given mass $M$ formed at a given constant rate \dm\ can be in equilibrium only if the values of $M$ and \dm\ are below these curves.
Between the two limits, the core is expected to have reached equilibrium, while the envelope remains dynamical, due to the slow sound speed in the coldest layers.
Above the upper limit, even the core has not converged to local equilibrium.
We see that these limits correspond approximately to the maximum accretion rates obtained in section~\ref{sec-res}.
The regime of full dynamical evolution can only be marginally reached with maximal accretion.
Notice, however, that these limits have been derived under the assumption of constant accretion rates,
that are above \dmax\ of equation~(\ref{eq-dmsms}) in the beginning of the run, when $M$ is small.
Interestingly, we see that the limit of dynamical accretion in the envelope corresponds approximately to the maximum rates for density profiles with $\rho/\robar=1\%$.
This density contrast is typical of the outer layers of rapidly accreting SMSs \citep{haemmerle2018a}.
Thus, dynamical evolution in the envelope would require a larger $\rho/\robar$ outside of the accretion shock than in the envelope,
i.e. a local density inversion near the stellar surface.
The coincidence between these various limits shows that inflow rates, which would prevent SMSs to evolve in equilibrium,
are hardly produced by gravitational collapse alone.
Reaching the regime of dynamical accretion implies necessarily a strong ram pressure at the accretion shock.

\subsection{GR instability and the maximum mass of SMSs}

Figure~\ref{fig-mmdot} indicates also the limit to equilibrium arising from GR instability in the case of spherical, non-rotating SMSs accreting at constant rates.
This constraint has been derived in \cite{haemmerle2020c,haemmerle2021a} from the hydrostatic \gva\ models of \cite{haemmerle2018a,haemmerle2019c},
which follow the evolution up to the instability only for rates 1 -- 10 \Mpy.
The final mass for the other rates is estimated by extrapolation of the tracks \citep{haemmerle2020c}.
The intersection between the limit in $M$ given by the GR instability and the one in \dm\ obtained in the present work gives a maximum value for both quantities.
In the absence of models at the largest rates, this intersection can be estimated only by extrapolation in $M$.
Figure~\ref{fig-mmdot} suggests an upper limit for the rate around $10^6$ \Mpy, and a maximum mass of $10^6-10^7$ \Ms,
in the optimal cases of homogeneous or $\rho\propto r^{-2}$ profiles.
For a density contrast $\rho/\robar=1\%$, the maximum permitted rate remains below $10^4$ \Mpy, and the mass hardly exceeds $10^6$ \Ms.

The impact of rotation on the GR instability in rapidly accreting SMSs has been addressed in \cite{haemmerle2021b}.
SMSs accreting at rates of atomically cooled haloes are expected to increase their final mass by a factor of a few because of rotation,
and to remain always $<10^6$ \Ms.
Rates $\gtrsim100$ \Mpy\ are required to exceed this threshold, and in this case final masses as high as $10^8-10^9$ \Ms\ could be reached.
Figure~\ref{fig-mmdot} shows that these rates are accessible before the GR instability is reached,
even without rotation and for the steep density profile $\rho/\robar=1\%$.
Notice that the rotation velocities required to reach masses $\sim10^9$ \Ms\ represent always $<2\%$ of the Keplerian velocity.
Thus, even in this case, the assumption of spherical symmetry, used to derive the maximum rate~(\ref{eq-dmsms}), remains relevent,
and the centrifugal barrier only weakly influences our estimates.

\subsection{Jeans instability and maximum accretion rate}

The expression of the maximum rate in equation~(\ref{eq-dmax}), as a function of the free-fall velocity,
takes a similar form as the isothermal rate of \cite{shu1977}, given instead by the sound speed $\vvvson$ as
\begin{equation}
\dm\sim{\vvvson\over G}\;.
\label{eq-shu}\end{equation}
This rate follows naturally from the free-fall of isolated Jeans masses $\dm\sim M_J/\tff$.
For systems with $M\gg M_J$, larger rates are found,
that scale roughly linearly with the number of Jeans masses contained in the cloud \citep{girichidis2011}.
The collapse of atomically cooled halos is set by the Jeans instability,
and the typical inflows found in hydrodynamical simulations are consistent with the thermal rate of equation~(\ref{eq-shu}).
In the absence of molecular hydrogen, primordial gas is thought to reach temperatures of $\sim10^4$ K,
which corresponds to a sound speed of $\sim10$ km s$^{-1}$ and a typical rate of 1 \Mpy\ \citep{latif2013e}.
In contrast, the large inflows found in simulations of galaxy mergers \citep{mayer2010,mayer2015,mayer2019}
rely on the dynamics of the merger, and are not set by the Jeans instability.
The most massive galaxies at redshift 8 -- 10 ($\sim10^{12}$~\Ms) are typically 4 -- 5 orders of magnitude more massive than primordial mini-haloes
($\sim10^{7-8}$~\Ms), but have similar Jeans masses $M_J$.
If the rate scales linearly with $M_J$, we naturally obtain accretion rates of $10^4$~\Mpy and above.
The hydrodynamical simulations show highly supersonic inflows at the centre of the potential well during the merger, which illustrates the fact that $\vff\gg\vson$,
and implies that the free-fall rate (\ref{eq-dmax}) of such configurations exceeds the thermal rate (\ref{eq-shu}).
The sound speed reflects the thermal content of the collapsing gas, that sets the Jeans mass,
while the maximum rate of equation~(\ref{eq-dmax}) is expressed only in terms of gravitational quantities,
without any assumption on the thermodynamics.
Thus, it represents an absolute limit that cannot be exceeded by gravity.

\section{Summary and conclusions}
\label{sec-out}

We have used the equation of continuity to derive maximum accretion rates allowed by pure gravitational collapse,
in which the conditions $\vit\leq\vff$ and $\rho\leq\robar$ are always satisfied.
These maximum allowed accretion rates are uniquely determined by the compactness (\ref{eq-rsr}) of the accretor.
If the mass-radius relation of the accretor is known, this maximum accretion rate is directly given by the mass of the accretor.

With the mass-radius relation~(\ref{eq-rsms}) of rapidly accreting SMSs,
we estimated the maximum permitted accretion rates as a function of their mass $M$ ($\dm\propto M^{3/4}$, see equation~\ref{eq-dmsms}).
The accretion rates 0.1 -- 10 \Mpy\ of atomically cooled haloes are consistent with gravitational infall once the central mass exceeds 1 \Ms.
Larger rates can be reached only once the star is massive ($M\gtrsim10$ \Ms).
Rates as large as $10^4-10^5$ \Mpy\ can only be achieved once the accretor has become supermassive ($M\gtrsim10^4$ \Ms).
For stars following such maximal accretion, we estimate the GR instability to be reached at masses $M\sim10^6-10^7$ \Ms\ in the non-rotating case,
up to $M\sim10^8-10^9$ \Ms\ with rotation.
At these masses, the instability is expected to lead to the direct formation of a supermassive black hole even in the case of metal-rich chemical composition,
in agreement with the galaxy merger scenario for direct collapse.

\begin{acknowledgements}
LH has received funding from the European Research Council (ERC) under the European Union's Horizon 2020 research and innovation programme (grant agreement No 833925, project STAREX). RSK acknowledges financial support from the German Research Foundation (DFG) via the collaborative research center (SFB 881, Project-ID 138713538) “The Milky Way System” (subprojects A1, B1, B2, and B8),  from the Heidelberg Cluster of Excellence ``STRUCTURES'' in the framework of Germany’s Excellence Strategy (grant EXC-2181/1, Project-ID 390900948), and from the ERC via the ERC Synergy Grant ``ECOGAL'' (grant 855130).
\end{acknowledgements}

\bibliographystyle{aa}
\bibliography{bib}

\end{document}